\documentclass[11pt,twoside]{article}

\usepackage{times,pifont}

\usepackage{euscript}
\usepackage{amsmath}	
\usepackage{amssymb,amsbsy}	
\usepackage[dvips]{epsfig}

\renewcommand{\baselinestretch}{1.1}
\newcount\cchap
\newcommand\fschapter[2]{%
\cleardoublepage
\vspace*{.5cm}
\advance\cchap by1
\begin{center}
\Large\sf #1\\[1ex]
\large\it #2
\end{center}
\bigskip
\addcontentsline{inh}{section}{\protect{#1}}
\setcounter{section}{0}
\setcounter{thm}{0} 
}

\makeatletter
\renewcommand{\section}{\@startsection
	{section}%
	{1}%
	{0mm}%
	{-2\baselineskip}%
	{\baselineskip}%
	{\normalfont\normalsize\bf}}
\makeatother

\makeatletter
\def\@begintheorem#1#2{\parskip0pt plus3pt\it \trivlist \item[\hskip 
\labelsep{\bf #1\ #2.}]}
\def\@opargbegintheorem#1#2#3{\parskip0pt plus3pt\it \trivlist
      \item[\hskip \labelsep{\bf #1\ #2\ #3.}]}
\makeatother

\makeatletter

\def\newdefinition#1{\@ifnextchar[{\@odefi{#1}}{\@ndefi{#1}}}

\def\@ndefi#1#2{%
\@ifnextchar[{\@xndefi{#1}{#2}}{\@yndefi{#1}{#2}}}

\def\@xndefi#1#2[#3]{\expandafter\@ifdefinable\csname #1\endcsname
{\@definecounter{#1}\@addtoreset{#1}{#3}%
\expandafter\xdef\csname the#1\endcsname{\expandafter\noexpand
  \csname the#3\endcsname \@deficountersep \@deficounter{#1}}%
\global\@namedef{#1}{\@defi{#1}{#2}}\global\@namedef{end#1}{\@enddefi}}}

\def\@yndefi#1#2{\expandafter\@ifdefinable\csname #1\endcsname
{\@definecounter{#1}%
\expandafter\xdef\csname the#1\endcsname{\@deficounter{#1}}%
\global\@namedef{#1}{\@defi{#1}{#2}}\global\@namedef{end#1}{\@enddefi}}}

\def\@odefi#1[#2]#3{\expandafter\@ifdefinable\csname #1\endcsname
  {\global\@namedef{the#1}{\@nameuse{the#2}}%
\global\@namedef{#1}{\@defi{#2}{#3}}%
\global\@namedef{end#1}{\@enddefi}}}

\def\@defi#1#2{\refstepcounter
    {#1}\@ifnextchar[{\@ydefi{#1}{#2}}{\@xdefi{#1}{#2}}}

\def\@xdefi#1#2{\@begindefi{#2}{\csname the#1\endcsname}\ignorespaces}
\def\@ydefi#1#2[#3]{\@opargbegindefi{#2}{\csname
       the#1\endcsname}{#3}\ignorespaces}

\def\@deficounter#1{\noexpand\arabic{#1}}
\def\@deficountersep{.}
\def\@begindefi#1#2{\it \trivlist \item[\hskip \labelsep{\bf #1\ #2.}]\rm}
\def\@opargbegindefi#1#2#3{\it \trivlist
      \item[\hskip \labelsep{\bf #1\ #2\ #3.}]\rm}
\def\@enddefi{\endtrivlist}

\makeatother

\newtheorem{thm}{Theorem}[section]
\newtheorem{coro}[thm]{Corollary}
\newtheorem{lemma}[thm]{Lemma}
\newtheorem{propo}[thm]{Proposition}

\newdefinition{defi}[thm]{Definition}

\newcommand\qedsymbol{\ding{113}}
\def\endproof{{\unskip\nobreak\hfil\penalty50
		      \hskip1em\hbox{}\nobreak\qedsymbol
		      \parfillskip=0pt\par\endtrivlist\addpenalty{-100}}} 
\def\proof{\trivlist
	\item[\hskip \labelsep{\it Proof.}]}

\newcommand\sw{\preceq}                             
\newcommand\op[1]{\langle #1\rangle}                
\newcommand\eps{\epsilon}                           
\newcommand\rel[1]{\stackrel{#1}{\Longrightarrow}}    
\renewcommand\L[1]{{\cal L}_{#1}}                   
\newcommand\C{{\cal C}}                             

\newcommand\NP{{\rm NP}}
\renewcommand\P{{\rm P}}
\newcommand\Sigp[1]{\Sigma^{\rm p}_{#1}}
\newcommand\Pip[1]{\Pi^{\rm p}_{#1}}
\newcommand\leafp{{\rm Leaf}^{\rm P}}

\newcommand\co{\mathord{\mbox{\rm co}}}

\newcommand\pari{\mathop{\oplus}}

\newcommand\N{{\mathbb{N}}}	
\newcommand\R{{\mathbb{R}}}	
\newcommand\set[2]{\bigl\{\,#1\bigm| #2\,\bigr\}}
\newcommand\seq{\subseteq}

\newcommand\eqdef{=_{\rm def}}
\newcommand{\df}{{\rm def}}
\renewcommand{\iff}{\mbox{$\;\Longleftrightarrow\;$}}

\begin{document}

%

\title{The Boolean Hierarchy over Level ${1/2}$ 
of the Straubing-Th\'erien Hierarchy}
\author{Heinz Schmitz \qquad Klaus W.~Wagner\\
Theoretische Informatik\\Universit\"at W\"urzburg\\Am Exerzierplatz 3\\
97072 W\"urzburg\\
\{schmitz, wagner\}@informatik.uni-wuerzburg.de}
\date{}
\maketitle
 
%

\begin{abstract}
For some fixed alphabet $A$ with $|A|\geq 2$, a language $L\seq A^*$
is in the class $\L{1/2}$ of the Straubing-Th\'erien hierarchy
if and only if it can be expressed as a finite union of languages 
$A^*a_1A^*a_2A^*\cdots A^*a_nA^*$, where $a_i\in A$ and $n\geq 0$.
The class $\L{1}$ is defined as the boolean closure of $\L{1/2}$.
It is known that the classes $\L{1/2}$ and $\L{1}$ are decidable. 
We give a membership criterion for the single classes of the boolean 
hierarchy over $\L{1/2}$.
From this criterion we can conclude that this boolean hierarchy is proper
and that its classes are decidable.
In finite model theory the latter implies the decidability of the classes 
of the boolean hierarchy over the class $\Sigma_1$ of the ${\rm FO}[<]$-logic.
Moreover we prove a ``forbidden-pattern'' characterization
of $\L{1}$ of the type: $L \in \L{1}$ if and only if a certain pattern
does not appear in the transition graph of a deterministic finite automaton accepting $L$.
We discuss complexity theoretical consequences of our results.

\noindent{\bf Classification:} 
finite automata, concatenation hierarchies, boolean hierarchy, decidability
\end{abstract}

%

\section{Introduction}

We contribute to the theory of finite automata and regular languages, 
as well as to complexity theory.
Particularly we deal with starfree regular languages.
These are languages which are constructed from alphabet letters only by using
boolean operations together with concatenation.
Alternating these two kinds of operations in order to
distinguish between combinatorial and sequential aspects
leads to the definition of concatenation hierarchies 
that exhaust the class of starfree languages.

Prominent examples are the dot-depth hierarchy, first studied in \cite{cobro71},
and the Straubing-Th\'erien hierarchy \cite{str81, the81, str85}. Both are known to be strict
\cite{brokna78} and closely related to each other.
Most naturally arising questions concerning these hierarchies
are of major interest in different research areas since there are close
connections to finite model theory, theory of finite
semigroups, topology, boolean circuits and others.
For an overview or as a good starting point to this rich field of research
see e.g.~the articles \cite{bro76, pin96b, pin96a, tho96}.

In this paper we deal with the so-called Straubing-Th\'erien hierarchy.
Let $A$ be some finite alphabet with $|A|\geq 2$.
For a class $\C$ of languages over $A^*$ let ${\rm POL}(\C)$ be its
polynomial closure, i.e.~the class of languages $L$ that can be written
as a finite union of languages 
$L_0a_1L_1a_2L_2\cdots L_{n-1}a_nL_n$, where $a_i\in A$,
$L_i\in\C$ and $n\geq 0$.
Denote by ${\rm BC}(\C)$ its boolean closure, i.e.~the closure of $\C$ under finite
union, finite intersection and complementation.
Then the Straubing-Th\'erien hierarchy can be defined as the family
of classes $\L{n/2}$, where we define
$\L{0}\eqdef\{\emptyset,A^*\}$,
$\L{n+1/2}\eqdef{\rm POL}(\L{n})$, and 
$\L{n+1}\eqdef{\rm BC}(\L{n+1/2})$ for $n\geq0$
(notations are adopted from \cite{pinweil97}).
We will also consider the classes $\co\L{n+1/2}$, where
$\co\C\eqdef\set{\overline{L}}{L\in\C}$ for a class $\C$.
It was shown by M.~Arfi in \cite{arfi87, arfi91} that the classes $\L{n+1/2}$
(and $\co\L{n+1/2}$) are closed under intersection.
For a language $L\seq A^*$ and 
a minimal $n$ with $L\in\L{n/2}$ we say that $L$ 
{\em has level} $n/2$.

The connection between first-order logic and the class of starfree languages 
goes back to the work of McNaughton and Papert \cite{mcpa71}.
The Straubing-Th\'erien hierarchy is related to the first-order logic
${\rm FO}[<]$ having only the binary relation $<$ and unary relations for the alphabet 
symbols from $A$. Let $\Sigma_k$ be the subclass of ${\rm FO}[<]$ which is defined by 
at most $k-1$ quantifier alternations, starting with an existential quantifier.
It has been proved by W.~Thomas in \cite{tho82} (see also \cite{pepi86}) that
$\Sigma_k$ formulas describe just the $\L{k-1/2}$ languages and that the
boolean combinations of 
$\Sigma_k$ formulas describe just the $\L{k}$ languages.

Unfortunately one main question about the Straubing-Th\'erien hierarchy, 
namely the question of 
the decidability of its classes, appears to be extremely difficult, although 
a lot of effort via different approaches has been invested.
The decidability problem can be stated as follows: given 
some $n\geq 0$ and a regular language $L$ presented by a deterministic finite
automaton, decide whether or not $L$ has level $n/2$.
To our knowledge, only levels $0, 1/2, 1,$ and $3/2$ are known to be 
decidable (cf.~\cite{pinweil97}).

The purpose of this paper is to start with an exact analysis of what happens between 
level $1/2$ and level $1$. Since $\L{1} = {\rm BC}(\L{1/2})$ and since ${\rm BC}(\L{1/2})$
is just the union of the classes $\L{1/2}(k)$ of the boolean hierarchy
over $\L{1/2}$ we study these classes $\L{1/2}(k)$ and their decidability.

J.~Stern \cite{stern85} proved the following interesting characterization of the 
class $\L{1}$ (the class of piecewise testable languages over alphabet $A$): 
A language $L\seq A^*$ is in
$\L{1}$ if and only if there does not exist an infinite chain $w_1, w_2, w_3,
\ldots$ of words where $w_{i+1}$ is an extension of $w_{i}$ and 
$w_{i}\in L \Leftrightarrow w_{i+1}\not\in L$ for $i = 1,2,3,\ldots$.
Let $m^+(L)$ be the length of a maximal chain of this kind starting with
$w_{1}\in L$. Using a normal form theorem for classes of boolean hierarchies,
we prove that $L \in \L{1/2}(k)$ if and only if $m^+(L) < k$. Since the 
latter property can be decided for fixed $k$ with a nondeterministic logarithmic
space algorithm, we can also decide the membership problem for the classes $\L{1/2}(k)$ 
with a nondeterministic logarithmic space algorithm. 
Furthermore we show that the
measure $m^+(L)$ is computable with an exponential space algorithm. 
Another consequence of the above membership criterion for the classes $\L{1/2}(k)$ is
the fact that this boolean hierarchy is indeed proper.

As a second contribution we prove a ``forbidden-pattern'' characterization
of $\L{1}$ of the type: $L \in \L{1}$ if and only if a certain 
pattern (see Figure \ref{fig3}) does not appear in a deterministic finite automaton 
accepting $L$. Such characterizations were already known for the classes
$\L{1/2}$ and $\L{3/2}$ \cite{pinweil97}. Our characterization easily 
provides a nondeterministic logspace decision algorithm for $\L{1}$.

There is a close connection between concatenation hierarchies and complexity
classes, both related via the so-called leaf language approach to define complexity classes.
This approach was introduced in \cite{bcs92, ver93} and led to a number of 
interesting results (cf.~\cite{helascvowa93, jemcth94, buvo98, crhevowa98}).
In particular in \cite{buvo98} it was shown that taking the languages from
$\L{k-1/2}$ as leaf languages yields
exactly the $k$-th class of the polynomial time hierarchy.
In the last section we state a result of this type relating the boolean hierarchy over 
level $1/2$ of the Straubing-Th\'erien hierarchy to the boolean hierarchy over $\NP$.
A similar, but ineffective result concerning the boolean hierarchy over level $1/2$ 
of the dot-depth hierarchy was obtained in \cite{bokust98}. 
Here we can make use of our decision algorithm, which is not known for the case
of the dot-depth hierarchy. 

Finally we want to make a remark concerning our methods.
First we note that the normalform results we use for the classes of the 
boolean hierarchy over $\L{1/2}$ are valid also for the classes of the 
boolean hierarchy over every class $\L{n+1/2}$. This combined with
the ``forbidden-pattern'' technique could work to achieve similar 
structural and decidability results for every level of the
Straubing-Th\'erien hierarchy.

%

\section{Preliminaries} 
We consider languages over an arbitrary finite alphabet $A$ with $|A|\geq 2$.
For a class $\C$ of languages, let ${\rm BC}(\C)$ be the boolean closure 
of $\C$, i.e.~${\rm BC}(\C)$ is the smallest class containing $\C$
and being closed under union, intersection and complementation.
For a class $\C$ which is closed under union and intersection, the boolean 
hierarchy over $\C$ is the family of classes $\C(k)$ 
and $\co\C(k)$ with $k\geq 1$, where  $\C(k)$ can be defined (besides 
many other equivalent possibilities, cf.~\cite{ksw87,caguha+88}) as
\[ \C(k)=_\df 
\underbrace{\C\pari\C\pari\cdots\pari\C}_{k \mbox{\scriptsize~times}},\]
where $\C\oplus\C=_\df\{A\bigtriangleup B~|~A\in\C,~B\in\C\}$,
$\bigtriangleup$ denotes the symmetric set difference and
$\co\C\eqdef\set{\overline{L}}{L\in\C}$.

The following lemma states some well-known properties of the classes of the 
boolean hierarchy over $\C$. Their normal form characterization in
statements 3 and 4 
provides one of the other possibilities of their definition.
\begin{lemma} \label{nf}
Let $\C$ be a class of languages which is closed under union and 
intersection, and let $k \geq 1$.
\begin{enumerate}
\item ${\rm BC}(\C) = \bigcup_{k\geq 1}\C(k)$.
\item $\C(k) \cup \co\C(k) \subseteq \C(k+1) \cap \co\C(k+1)$.
\item $ L \in \C(2k-1)$ if and only if there exist languages 
      $L_1,L_2,\ldots,L_{2k-1} \in \C$ such that\\ 
      $L_1 \supseteq L_2 \supseteq \cdots \supseteq L_{2k-1}$ and
      $L = \bigcup_{i=1}^{k-1}(L_{2i-1}\backslash L_{2i}) \cup L_{2k-1}$.
\item $ L \in \C(2k)$ if and only if there exist languages
      $L_1,L_2,\ldots,L_{2k} \in \C$ such that\\ 
      $L_1 \supseteq L_2 \supseteq \cdots \supseteq L_{2k}$ and 
      $L = \bigcup_{i=1}^{k}(L_{2i-1}\backslash L_{2i})$.
\end{enumerate}
\end{lemma}

For a class $\C$ of languages, let ${\rm POL}(\C)$ be its
polynomial closure, i.e.~the class of languages $L$ that can be written
as a finite union of languages 
$L_0a_1L_1a_2L_2\cdots L_{n-1}a_nL_n$, where $a_i\in A$,
$L_i\in\C$ and $n\geq 0$.
Then the Straubing-Th\'erien hierarchy can be defined as the following
family of classes, where notations are adopted from \cite{pinweil97}. 
\begin{enumerate}
\item $\L{0}\eqdef\{\emptyset,A^*\}$
\item $\L{n+1/2}\eqdef{\rm POL}(\L{n})$ for $n\geq0$
\item $\L{n+1}\eqdef{\rm BC}(\L{n+1/2})$ for $n\geq0$
\end{enumerate}
We will also take into consideration the classes $\co\L{n+1/2}$.
Any class $\L{n+1/2}$ can be equivalently defined as the closure of the
class $\L{n}$ under union, intersection and the so-called marked 
concatenation (cf.~\cite{arfi87, arfi91}).
Consequently, the results of Lemma \ref{nf} apply also to the classes
${\cal C} = \L{n+1/2}$.
For a language $L\seq A^*$ and 
a minimal $n$ with $L\in\L{n/2}$ we say that
$L$ {\em has level} $n/2$.

Next we point out a very natural connection between the 
Straubing-Th\'erien hierarchy and a certain logic over finite words.
We define formulas using the binary relation symbol $<$ and
unary relation symbols $\pi_a$ for each letter $a\in A$. 
Atomic formulas are of the type $x<y$, $x=y$ and $\pi_ax$, with variables $x,y$. 
Then formulas are contructed from atomic formulas by using the connectives
$\neg, \vee, \wedge$ and quatifiers $\exists, \forall$ bounding variables.
Let $\Sigma_k$ ($\Pi_k$) be the subclass of such formulas which have 
at most $k-1$ quantifier alternations, starting with an existential (universal, resp.)
quantifier.
We say a language $L\seq A^*$ is ${\rm FO}[<]$-definable if there exists a
sentence $\phi$ (i.e.~a formula of the above type without free variables)
such that all words $w\in L$ satisfy $\phi$
when variables are interpreted as positions in $w$, $\pi_ax$ means the letter
at position $x$ is $a$, and $<$ is the usual $<$-relation on $\{1,\ldots,|w|\}$.

\begin{thm}[\cite{tho82, pepi86}] \label{logic}
Let $k \geq 1$ , and let $L\seq A^*$ be any language.
\begin{enumerate}
\item $L \in \L{k-1/2}$ if and only if $L$ is ${\rm FO}[<]$-definable by a $\Sigma_k$
      formula.
\item $L \in \co\L{k-1/2}$ if and only if $L$ is ${\rm FO}[<]$-definable by a $\Pi_k$
      formula.
\item $L \in \L{k}$ if and only if $L$ is ${\rm FO}[<]$-definable by a boolean
      combination of $\Sigma_k$ formulas.
\end{enumerate}
\end{thm}

Let $\eps$ be the empty word.
We denote by $\sw$ the subword relation on $A^*$, i.e.
$w\sw v$ if and only if there exist $n \geq 1$, 
$a_1,a_2,\ldots,a_n \in A$ and $v_0,v_1,\ldots,v_n \in A^*$ such that
$w=a_1a_2\cdots a_n$ and $v=v_0a_1v_1a_2v_2\cdots a_nv_n$.
For $w \in A^*$ we define 
$\op{w}_\sw \eqdef \{v~|~w\sw v\}$
as the set of all words having $w$ as a subword, i.e.~
$\op{a_1a_2\cdots a_n}_\sw = A^*a_1A^*a_2A^*\cdots A^*a_nA^*$
for all  $n \geq 1$ and $a_1,a_2,\ldots,a_n \in A$.
Moreover, for a language L let $\op{L}_\sw\eqdef\bigcup_{w\in L} \op{w}_\sw$
be the set of all words having a subword in $L$.
For a word $w=a_1a_2\cdots a_n$ we denote with $w^R$ its reverse, 
i.e.~$w^R\eqdef a_na_{n-1}\cdots a_1$, and for a language $L$ let
$L^R\eqdef\set{w^R}{w\in L}$.
We will denote infinite sequences of words $\{w_i\}_{i=1}^{\infty}$
for short as $\{w_i\}$.

As is standard, a deterministic finite automaton (dfa) $F$ is given by 
$F=(A, S, \delta, s_0, S^{'})$, 
where $A$ is its input alphabet, $S$ is its set of states, 
$\delta:A\times S \rightarrow S $ is its
transition function, $s_0\in S$ is the starting state and $S^{'}\seq S$ is the
set of accepting states.
We consider nondeterministic finite automata (nfa) as well, where
$\delta:A\times S \rightarrow  2^S$. 
With $L(F)$ we denote the language accepted by an automaton $F$.
As usual we extend transition functions to input words, and we denote by $|F|$ 
the number of states of $F$.

\begin{thm}\label{sigma-open}
For every $L\seq A^*$ the following are equivalent:
\begin{enumerate}
\item[(1)] $L\in \L{1/2}$
\item[(2)] $L$ is a finite union of sets $\op{w}_\sw$
\item[(3)] $L$ is regular and $\op{L}_\sw = L$
\end{enumerate}
\end{thm}
 
\proof
The equivalence (1) $\Leftrightarrow$ (2) is by definition,
and (2) $\Rightarrow$ (3) is obvious.
For (3) $\Rightarrow$ (2), let $F$ be a dfa such that 
$\op{L(F)}_\sw = L(F)$. Let $F'$ be the nfa which is constructed 
from $F$ be introducing for every state  and every $a \in A$ a
simple loop with $a$. Obviously, $L(F') = L(F)$. Now convert $F'$
into the nfa $F''$ by removing all nontrivial loops, i.e.~by
keeping only the paths leading directly from the starting state to an 
accepting state. Also, $L(F'') = L(F')$. Now, $L(F'')$ is the union
of all $\op{a_1a_2\cdots a_n}_\sw$ where $a_1a_2\cdots a_n$ is
a path in $F''$ leading directly from the starting state to an 
accepting state.
\endproof

We assume the reader to be familiar with complexity classes of common interest
such as ${\rm NL}$, $\P$, $\NP$ and the levels $\Sigp{k}$ of the polynomial time hierarchy.

%

\section{Alternating Word Extension Chains}

We will obtain a membership criterion for the classes $\L{1/2}(k)$
by examining the number of alternations that may occur in
a sequence of words, where each word is an extension of its predecessor. 
Let us first make this notion precise.

\begin{defi}[\cite{stern85}]
Let $L\seq A^*$, $m\geq 0$ and $w,v\in A^*$.
We say that {\em  $v$ is reachable from $w$ by an $m$-alternating 
word extension chain with respect to $L$},
i.s.~$w \rel{m}_L v$, 
if and only if there exist
$w_0, w_1,\ldots, w_m\in A^*$ such that
\begin{enumerate}
\item $w=w_0\sw w_1 \sw w_2 \sw \ldots \sw w_m \sw v$, and
\item $w_i\in L$ if and only if $w_{i+1}\not\in L$ for $1\leq i\leq m-1$.
\end{enumerate}
\end{defi}

Next we take a closer look at such chains and define 
the sets of words that can be reached from a word (not) in
a given language $L$ by at least $m$ alternations.

\begin{defi}
For a language $L\seq A^*$ and $m\geq 0$ we define
\begin{enumerate}
\item $L^+(m)\eqdef\set{v\in A^*}{\exists w~(w\in L \wedge w \rel{m}_L v)}.$
\item $L^-(m)\eqdef\set{v\in A^*}{\exists w~(w\not\in L \wedge w \rel{m}_L v)}.$
\end{enumerate}
\end{defi}

We summarize some properties of $L^+(m)$ and $L^-(m)$ in the following 
proposition.

\begin{propo} \label{prop1}
For a language $L$ and $m \geq 0$ the following statements hold:
\begin{enumerate}
\item $L^-(m)=\overline{L}^+(m)$.
\item $L^+(0) = \op{L}_\sw$ and  $L^-(0) = \op{\overline{L}}_\sw$.
\item $L^+(m+1)\cup L^-(m+1) \subseteq L^+(m)\cap L^-(m)$.
\item $v\not\in L^+(m) \cup L^-(m)$ for all $m > |v|$.
\item $\bigcap_{m\geq 0}L^+(m) = \bigcap_{m\geq 0}L^-(m) = \emptyset$.
\item $L^+(m)\not=\emptyset$ implies $L^+(m+1)\subsetneq L^+(m)$, and
      $L^-(m)\not=\emptyset$ implies $L^-(m+1)\subsetneq L^-(m)$.
\item $L^+(m)=\op{L^+(m)}_\sw$ and $L^-(m)=\op{L^-(m)}_\sw$. 
\end{enumerate}
\end{propo}
 
Now we show that 
any language $L$ can be expressed as a possibly infinite union
of set differences of sets $L^+(m)$ and $L^-(m)$.

\begin{propo} \label{prop3}
\renewcommand{\baselinestretch}{1.5}
\small\normalsize
For a language $L\seq A^*$ the following statements hold:
\begin{enumerate}
\item $L=\bigcup_{m\geq 0}^{\infty}\big(L^+(2m)\backslash L^+(2m+1)\big)$
      and \\
      $\overline{L}=\big(A^*\backslash L^+(0)\big)\cup
        \bigcup_{m\geq 1}^{\infty}\big(L^+(2m-1)\backslash L^+(2m)\big).$
\item $\overline{L}=
        \bigcup_{m\geq 0}^{\infty}\big(L^-(2m)\backslash L^-(2m+1)\big)$
      and\\ 
      $L=\big(A^*\backslash L^-(0)\big)\cup
        \bigcup_{m\geq 1}^{\infty}\big(L^-(2m-1)\backslash L^-(2m)\big).$

\end{enumerate}
\end{propo}

\proof
Let $m\geq 0$ and $v\in L^+(2m)\backslash L^+(2m+1)$.
Because of $v\in L^+(2m)$ there exists a $w\in L$ with $w \rel{2m}_L v$.
Now observe that if $v\not\in L$ then $w \rel{2m+1}_L v$ witnessed by the
same word extension chain as before, which is a contradiction to
$v\not\in L^+(2m+1)$. Hence $v \in L$.

In the same way one proves that $v\in L^+(2m-1)\backslash L^+(2m)$
implies $v \not\in L$ for $m\geq 1$, and that $v\in A^*\backslash L^+(0)$
implies $v \not\in L$.

Statement 2 follows from 1 by Proposition \ref{prop1}.1.
\endproof

Now we want to show that for a regular set $L$ the sets $L^+(m)$ and 
$L^-(m)$ belong to $\L{1/2}$.
Proposition \ref{prop1}.7 already says that 
$L^+(m)=\op{L^+(m)}_\sw$ and $L^-(m)=\op{L^-(m)}_\sw$. With 
Theorem \ref{sigma-open} it remains to show that they are regular.

\begin{lemma} \label{lemma1}
If $L\seq A^*$ is regular and $m\geq 0$, then $L^+(m)$ and $L^-(m)$
are regular as well.
\end{lemma}

\proof
Let $F=(A, S, \delta, s_0, S^{'})$ be a deterministic finite automaton 
accepting $L$.
We construct a nondeterministic finite automaton $F_m$ that accepts 
$L^+(m)$ and that realizes the
idea of guessing a $m$-alternating chain of subwords of the input. 
Define $F_m\eqdef(A, S_m , \delta_m, s_0^m, S^{'}_m)$ as
\begin{itemize}
\item $S_m\eqdef
      \underbrace{S\times S\times\cdots \times S}_{(m+1){\rm -times}}$
\item $s_0^m\eqdef(s_0, s_0, \ldots , s_0)$
\item $\delta_m((s_1, s_2, \ldots , s_{m+1}),a)\eqdef\\
      \hspace*{40mm}
      \set{(s_1, s_2, \ldots, s_i, \delta(s_{i+1},a), \ldots, 
                                        \delta(s_{m+1},a))}{0\leq i\leq m+1}$
\item $S^{'}_m\eqdef \set{(s_1, s_2, \ldots , s_{m+1})}
                         {(s_{i}\in S' \leftrightarrow i \mbox{ odd}) 
                  \mbox{ for } i = 1,\ldots ,m+1}$.
\end{itemize}

We observe that $(s_1, s_2, \ldots , s_{m+1})\in\delta_m(s_0^m,v)$ if and 
only if there exist words $w_1,\ldots, w_{m+1}\in A^*$ such that
$w_1 \sw w_2 \sw \ldots \sw w_{m+1} \sw v$ and 
$\delta(s_0,w_i)=s_i$ for $i = 1,\ldots ,m+1$.

Now we can conclude:
\[ v\in L(F_m) \begin{array}[t]{cl}
   \iff & \delta_m(s_0^m,v)\cap S^{'}_m \not= \emptyset \\
   \iff & \mbox{there exist~} s_1, s_2, \ldots , s_{m+1} \mbox{~such that~}
          (s_1, s_2, \ldots , s_{m+1})\in\delta_m(s_0^m,v)\cap S^{'}_m \\
   \iff & \mbox{there exist~} w_1,\ldots, w_{m+1} \mbox{~such that~} 
          w_1 \sw w_2 \sw \ldots \sw w_{m+1} \sw v \mbox{~and~} \\
   ~    & (\delta(s_0,w_{i})\in S^{'} \leftrightarrow i \mbox{ odd})
          \mbox{~for~} i = 1,\ldots ,m+1\\
   \iff & v\in L^+(m)
          \end{array}
\]

Because of $L^-(m) = \overline{L}^+(m)$ we obtain that $L^-(m)$ is
also regular.
\endproof

\begin{coro} \label{coro1}
If $L\seq A^*$ is regular and $m\geq 0$, then $L^+(m)$ and $L^-(m)$ are in $\L{1/2}$.
\end{coro}

In order to measure the number of inevitable alternations that occur
with respect to a given language $L$ we look for the maximal $m$ such
that the sets $L^+(m)$ and $L^-(m)$ are not empty.

\begin{defi}
For a language $L\seq A^*$ we set $m^+(L)\eqdef\max\set{m}{L^+(m)\not=\emptyset}$
and $m^-(L)\eqdef\max\set{m}{L^-(m)\not=\emptyset}$.
\end{defi}

The following proposition is an immediate consequence of Proposition 
\ref{prop1}.

\begin{propo} \label{prop4}
For any language $L\seq A^*$ it holds that 
\begin{enumerate} 
\item $m^+(L)=\infty$~if and only if~$m^-(L)=\infty$,
\item if~$m^+(L)<\infty$~then~$|m^+(L)-m^-(L)|=1$, and
\item $m^+(L)=m^-(\overline{L})$.
\end{enumerate}
\end{propo}

%

\section{A Criterion for Membership in $\L{1/2}(k)$}

The measure $m^+$ has already been used by J.~Stern to characterize  
$\L{1}={\rm BC}(\L{1/2})$, i.e.~the piecewise testable languages over alphabet $A$.

\begin{thm}[\cite{stern85}] \label{stern}
A language $L\seq A^*$ belongs to $\L{1}$ if and only if $m^+(L)$ is finite.
\end{thm}

Now we will relate the single classes of the boolean hierarchy over 
$\L{1/2}$ to particular values of $m^+$ and $m^-$. This theorem 
then has the preceding one as a corollary.
\begin{thm}\label{Sk-mk}
Let $L\seq A^*$ be a language and $k\geq 1$.
\begin{enumerate}
\item $L\in\L{1/2}(k)$ if and only if $L$ is regular and $m^+(L)< k$.
\item $L\in\co\L{1/2}(k)$ if and only if $L$ is regular and $m^-(L)< k$.
\end{enumerate}
\end{thm}

\proof
We prove Statement 1; Statement 2 then follows immediately by 
Proposition \ref{prop4}.3.
We restrict ourselves to the case of even $k$, the other case being proved 
analogously.

Let $L$ be regular and $m^+(L)< 2k$. Then $L^+(i)=\emptyset$ for all 
$i \geq 2k$.
By Proposition \ref{prop3}.1 we can write $L$ as 
 $$L=\bigcup_{i = 0}^{k-1}\big(L^-(2i)\backslash L^-(2i+1)\big),$$
and Corollary \ref{coro1} shows that we can use Lemma \ref{nf}.4
to obtain $L \in \L{1/2}(2k)$.

Now suppose $L\in\L{1/2}(2k)$. Then $L$ is regular and 
again by Lemma \ref{nf}.4 there exist languages  
$L_1,L_2,\ldots,L_{2k} \in \L{1/2}$ such that
$L_1 \supseteq L_2 \supseteq \cdots \supseteq L_{2k}$ and
$L = \bigcup_{i=1}^{k}(L_{2i-1}\backslash L_{2i})$.
Setting $L_0\eqdef A^*$ and $L_{2k+1}\eqdef\emptyset$ we obtain
$\overline{L} = \bigcup_{i=0}^{k}(L_{2i}\backslash L_{2i+1})$.

Assume that $L^+(2k)\not=\emptyset$. 
Then by definition of $L^+(2k)$ there exist $w\in L$, some $v\in A^*$ and 
$w_0, w_1,\ldots , w_{2k}\in A^*$ such that
$w=w_0\sw w_1 \sw w_2 \sw \ldots \sw w_{2k} \sw v$ with $w_{2i}\in L$ and
$w_{2i-1}\not\in L$.
For any $i\in\{0,1,\ldots , 2k-1\}$ there must be two indices 
$j,j'\in \{0, \ldots ,2k\}$ with $w_i\in L_j\backslash L_{j+1}$ and
$w_{i+1}\in L_{j'}\backslash L_{j'+1}$.
Since $w_{i}\in L \Leftrightarrow w_{i+1}\not\in L$ these indices
must be different.
Note with Theorem \ref{sigma-open} that $\op{L_j}_\sw = L_j$ for all $j$. 
So from $w_i\sw w_{i+1}$ we can conclude that
$w_{i+1}\in L_{j}$ as well, which implies $j' > j$.
Consequently, the words $w_0, w_1,\ldots , w_{2k}$ are
in $2k+1$ different sets $L_j\backslash L_{j+1}$ with
$j\geq 1$ (since $w_0 \in L \subseteq L_1$).
This is a contradiction since there are only $2k$ such sets. 
Hence $m^+(L)< 2k$.
\endproof

In the remainder of this section we will give two applications of the above
criterion for membership in $\L{1/2}(k)$.
First, we can conclude that the boolean hierarchy over $\L{1/2}$ is a proper
hierarchy.

\begin{thm}
For every $k\geq 1$, 
\[
 \L{1/2}(k)\subsetneq\L{1/2}(k+1).
\]
\end{thm}

\proof
Fix some $a\in A$, and define $|w|_a$ to be the number of occurences of $a$ in $w\in A^*$.
For $k\geq 1$ define
\begin{enumerate} 
 \item $M_{2k-1}\eqdef\set{w\in A^*}{|w|_a \mbox{~is~odd~or~}|w|_a > 2k-1}$, and 
 \item $M_{2k}\eqdef\set{w\in A^*}{|w|_a \mbox{~is~odd~and~}|w|_a\leq 2k}$.
\end{enumerate}
Obviously it holds that $m^-(M_k)=k$ and $m^+(M_k)=k-1$.
By Theorem \ref{Sk-mk} we obtain $M_k\in\L{1/2}(k)\backslash\co\L{1/2}(k)$, and
by Lemma \ref{nf}.2 we get $\L{1/2}(k)\subsetneq\L{1/2}(k+1)$.
\endproof

Next we consider the decidability of the classes $\L{1/2}(k)$.
For a given dfa $F$, the equivalence 
$L(F)\in\L{1/2}(k) \Leftrightarrow m^+(L(F))< k$ given by Theorem \ref{Sk-mk}
can be used to obtain a decision procedure for 
the question $L(F)\stackrel{?}{\in}\L{1/2}(k)$. 
This follows from the next lemma.
Here and in the sequel we assume that a regular language is given
by a deterministic finite automaton.

\begin{lemma} \label{decide}
Given a dfa $F$ and $k\geq 1$, the questions $m^+(L(F))\stackrel{?}<k$ and
$m^-(L(F))\stackrel{?}<k$ are decidable in nondeterministic space $k\cdot\log |F|$. 
\end{lemma}

\proof
Note that $m^+(L(F))<k \Leftrightarrow L(F)^+(k)=\emptyset  
\Leftrightarrow L(F_k)=\emptyset$ where $F_k$ is the nfa constructed in the 
proof of Lemma \ref{lemma1}. Obviously,  $L(F_k)=\emptyset$ is equivalent
with the non-existence of a path between the starting state of $F_k$ and
one of its accepting states. Hence, we have to solve the graph 
non-accessibility problem for the transition graph of $F_k$ which is of 
size $|A|\cdot|F|^{k+1}$. This can be done
in co-nondeterminstic space $\log(|F|^{k+1})= (k+1)\cdot\log|F|$ which is the 
same as nondeterministic space $k\cdot\log |F|$ \cite{im88, sze87}.
\endproof

\begin{thm}
For fixed $k\geq 1$,
the decision problems for $\L{1/2}(k)$ and $\co\L{1/2}(k)$ are in {\em NL}. 
\end{thm}

We are able to decide the question  $m^+(L(F))\stackrel{?}<k$ for given dfa
$F$ and $k \geq 1$. However, this does not mean automatically that we are 
able to compute $m^+(L(F))$ effectively. That this is indeed possible
can be concluded from the following dichotomy-lemma by J.~Stern.

\begin{lemma}[\cite{stern85}]\label{stern-dicho}
For a deterministic finite automaton $F$, 
 $$m^+(L(F)) < \infty \Longleftrightarrow m^+(L(F)) \leq 2^{|F|\cdot|A|^2}.$$
\end{lemma}

This dichotomy enables us to compute the measure $m^+(L(F))$ 
simply by deciding the questions $m^+(L(F))\stackrel{?}<k$ for 
$k=1,2,\ldots ,2^{|F|\cdot|A|^2}+1$  with help of Lemma \ref{decide}.

\begin{thm} \label{compm}
The measures $m^+(L)$ and $m^-(L)$ for a regular language $L$ are computable in 
space $2^{{\cal O}(|F|)}$.
\end{thm}

Due to the close connection to the ${\rm FO}[<]$-logic 
(Theorem \ref{logic}) we immediately have the following corollary.

\begin{coro}
The classes of the boolean hierarchy over the class $\Sigma_1$ of ${\rm FO}[<]$-logic
are decidable.
\end{coro}

%

\section{A Pattern Characterization for $\L{1}$}

In this section we give a ``forbidden-pattern'' characterization of the class
$\L{1}$ (for other characterizations of this class see \cite{simon75,stern85}).
First we define significant patterns that lead to infinite alternating extension chains. 
The technically involved part in the proof of the following theorem is 
to show conversely that an infinite alternating extension chain implies the 
occurence of such a pattern.
For this end 
we continuously select suitable infinite subchains of an infinite chain, 
we emphasize on the position in a word where insertion of a letter leads to alternation and
we extensively exploit the finiteness of an automaton.

We say that the dfa  $F=(A, S, \delta, s_0, S^{'})$ has the pattern P$_1$
(cf. Figure \ref{fig1}) if there exist $v,x,y,z\in A^*, a\in A$ and states 
$s_1,s_2,s_3\in S$ such that 
      $ya\sw v$, $\delta(s_0,x)=\delta(s_1,v)=s_1$,
      $\delta(s_1,y)=s_2$, $\delta(s_2,a)=s_3$ and
      $\delta(s_2,z)\in S^{'} \Leftrightarrow \delta(s_3,z)\not\in S^{'}$.

We say that the dfa $F$ has the pattern P$_2$ (cf. Figure \ref{fig2}) if there exist 
$u,x,z,z'\in A^*, a\in A$ and states $s_1,s_2,s_3,s_4\in S$ such that 
      $az\sw u$, $\delta(s_0,x)=s_1$, $\delta(s_1,a)=s_2$,
      $\delta(s_1,z)=\delta(s_3,u)=s_3$, 
      $\delta(s_2,z)=\delta(s_4,u)=s_4$ and  
      $\delta(s_3,z')\in S^{'} \Leftrightarrow \delta(s_4,z')\not\in S^{'}$.

We say that the dfa  $F$ has the pattern P$_3$
(cf. Figure \ref{fig3}) if there exist $u,v,x,y,z,z'\in A^*, a\in A$ and states 
$s_1,s_2,s_3,s_4,s_5\in S$ such that 
      $ya\sw v$ or $az\sw u$, $\delta(s_0,x)=\delta(s_1,v)=s_1$,
      $\delta(s_1,y)=s_2$, $\delta(s_2,a)=s_3$,
      $\delta(s_2,z)=\delta(s_4,u)=s_4$,
      $\delta(s_3,z)=\delta(s_5,u)=s_5$ and 
      $\delta(s_4,z')\in S^{'} \Leftrightarrow \delta(s_5,z')\not\in S^{'}$.

\begin{figure}[t] 
\begin{center}
\input{pattern1.pstex_t}%
\caption{\label{fig1}Pattern P$_1$ with $ya\sw v$ and $s\in S^{'} \Leftrightarrow s'\not\in S^{'}$.}
 
\input{pattern2.pstex_t}%
\caption{\label{fig2}Pattern P$_2$ with $az\sw u$ and $s\in S^{'} \Leftrightarrow s'\not\in S^{'}$.}
 
\input{pattern3.pstex_t}%
\caption{\label{fig3}Pattern P$_3$ with $ya\sw v$ or $az\sw u$, 
         and $s\in S^{'} \Leftrightarrow s'\not\in S^{'}$.}
\end{center}
\end{figure}

\begin{thm}\label{pattern}
Let $F$ be a dfa and let $\hat{F}$ be a dfa such that $L(F)^R=L(\hat{F})$.
Then the following are equivalent:
\begin{enumerate}
\item[(1)] $L(F)\in \L{1}$,
\item[(2)] neither $F$ nor $\hat{F}$ does have the pattern {\em P}$_1$,
\item[(3)] neither $F$ nor $\hat{F}$ does have the pattern {\em P}$_2$,
\item[(4)] $F$ does not have the pattern {\em P}$_3$.
\end{enumerate}
\end{thm}

In the proof we will make use of the following easy to see lemma.
\begin{lemma} \label{sequence}
Let $\{\alpha_i\}$ be a sequence of real numbers such that 
$0<\alpha_i<1$ and $\alpha_i\not=\alpha_j$ for $i\not= j$.
Then there exists an infinite monotonic subsequence of $\{\alpha_i\}$.
\end{lemma} 

\trivlist \item[\hskip \labelsep{\it Proof of Theorem \ref{pattern}.}]
$(2)\Rightarrow (1)$: 
Assume that $L(F)\not\in \L{1}$ for some dfa $F=(A, S, \delta, s_0, S^{'})$. 
We have to show that $F$ has pattern P$_1$ or any $\hat{F}$ with $L(F)^R=L(\hat{F})$
has pattern P$_2$.
First we conclude with Theorem \ref{stern} that $m^+(L(F))$ is infinite and we can
assume w.l.o.g. that there exists an infinite sequence of words $\{w_j\}$ and a letter
$a\in A$ such that $w_j\sw w_{j+1}$ for all $j\geq 1$, and 
$w_{2i-1}=w_i^{'}w_i^{''}$, $w_{2i}=w_i^{'}aw_i^{''}$,
$\delta(s_0,w_{2i-1})\not\in S^{'}$ and $\delta(s_0,w_{2i})\in S^{'}$ for all $i\geq 1$.
Next we introduce markers $m_i$ at the positions where $a$ is inserted
when going from $w_{2i-1}$ to $w_{2i}$, i.e.~the word $w_i^{'}aw_i^{''}$ has
markers $m_1,m_2,\ldots,m_i$.
To show the existence of an infinite subsequence of words which is monotonic with
respect to the insertion positions of the letter $a$,
we inductively attach values $\alpha_i\in\R$ to each marker $m_i$ as follows:
Let $\alpha_{i+1}\eqdef(\beta_{i+1}+\gamma_{i+1})/2$ with 
$\beta_{i+1}\eqdef\max\big(\set{\alpha_j}
     {1\leq j \leq i \mbox{~and~marker~}m_j\mbox{~is~left~to~}m_{i+1}}\cup\{0\}\big)$ 
and
$\gamma_{i+1}\eqdef\min\big(\set{\alpha_j}
     {1\leq j \leq i \mbox{~and~marker~}m_j\mbox{~is~right~to~}m_{i+1}}\cup\{1\}\big)$.
We observe that $m_i$ is left to $m_j$ if and only if $\alpha_i<\alpha_j$.
Now Lemma \ref{sequence} tells us that there is an infinite strictly monotonic subsequence 
of $\{\alpha_i\}$. We distinguish two cases.

{\em Case 1}.
Assume that there exists an infinite strictly increasing subsequence of $\{\alpha_i\}$, i.e.~there
is a mapping $\tau:\N\to\N$ such that $\tau(i)<\tau(i+1)$ and $\alpha_{\tau(i)}<\alpha_{\tau(i+1)}$
for all $i\geq 1$.
For simplicity we redefine $w_i\eqdef w_{\tau(i)}$ and summarize the properties of the
sequence selected in this way.
For all $i\geq 1$ we have 
\begin{enumerate}
\item $w_{2i-1}=w_i^{'}w_i^{''}\sw w_i^{'}aw_i^{''}=w_{2i}\sw w_{2i+1}=w_{i+1}^{'}w_{i+1}^{''}$,
\item $w_i^{'}a \sw w_{i+1}^{'}$, 
\item $\delta(s_0,w_{2i-1})\not\in S^{'}$ and $\delta(s_0,w_{2i})\in S^{'}$.
\end{enumerate}
We use the sequence $\{w_i^{'}\}$ as a starting point for subsequent selections of
sequences $\{w_{i,k}\}$ for $k=0,1,2,\ldots$ 
all having the properties stated in the following claim. 
Using the finiteness of the set of states will then enable us to find 
the pattern P$_1$ in $F$.
In the following notations a superscript in combination with a subscript denotes an index.

{\em\bf Claim}.
For every $k\geq0$ there exists a state $s_k\in S$ and an infinite subsequence
$\{w_{i,k}\}$ of $\{w_i^{'}\}$ such that for all $k,i$ 
there are words $v^1_{i,k},v^2_{i,k},\ldots,v^k_{i,k},u_{i,k}\in A^*$ with
\begin{enumerate}
\item[a.] $w_{i,k}=v^1_{i,k}av^2_{i,k}a\cdots av^k_{i,k}au_{i,k}$,
\item[b.] $v^j_{i,k}\sw v^j_{i+1,k}$ for $1\leq j\leq k$,
\item[c.] $u_{1,k-1}\sw v^k_{i,k}$ for $k\geq 1$,
\item[d.] $u_{i,k}a\sw u_{i+1,k}$, and
\item[e.] $\delta(s_0,v^1_{i,k}av^2_{i,k}a\cdots av^j_{i,k}a)=s_j$ for $1\leq j\leq k$.
\end{enumerate}

\trivlist \item[\hskip \labelsep{\em Proof of claim.}]
We proceed by induction on $k$.
The case $k=0$ is easy to see with $w_{i,0}\eqdef u_{i,0}=w_i^{'}$.
Starting with $\{w_{i,k}\}$ we show how to select a subsequence $\{w_{i,k+1}\}$
fulfilling the assertions of the claim.
First we observe that we can conclude from $u_{i,k}a\sw u_{i+1,k}$ that 
$u_{1,k}a\sw u_{i,k}$ for all $i\geq 2$.
Now for every $i\geq 2$ we can identify in $u_{i,k}$ a word left (right, resp.)
of this particular letter $a$, i.e. there are words $v^{k+1}_{i,k}$ and $u^{'}_{i,k}$ 
such that $u_{i,k}=v^{k+1}_{i,k}au^{'}_{i,k}$, 
$u_{1,k}\sw v^{k+1}_{i,k}\sw v^{k+1}_{i+1,k}$ 
and $u^{'}_{i,k}a\sw u^{'}_{i+1,k}$.
Hence we can write each $w_{i,k}$ as 
$w_{i,k}=v^1_{i,k}av^2_{i,k}a\cdots av^k_{i,k}av^{k+1}_{i,k}au^{'}_{i,k}$.
Due to the finiteness of the set of states of $F$ we can conclude
that there exists a state $s_{k+1}\in S$ and a strictly increasing mapping $\tau:\N\to\N$ 
such that
$\delta(s_0,v^1_{\tau(i),k}av^2_{\tau(i),k}a\cdots av^k_{\tau(i),k}av^{k+1}_{\tau(i),k}a)=s_{k+1}$.
Now we define 
$w_{i,k+1}\eqdef w_{\tau(i),k}$, $v^j_{i,k+1}\eqdef v^j_{\tau(i),k}$ for $1\leq j\leq k+1$ and
$u_{i,k+1}\eqdef u^{'}_{\tau(i),k}$.
We leave the verification of the assertions a to e for $\{w_{i,k+1}\}$
as an exercise.
{\unskip\nobreak\hfil\penalty50\hskip1em\hbox{}\nobreak{\em (End proof of claim)}
                      \parfillskip=0pt\par\endtrivlist\addpenalty{-100}}
We keep the notations of the claim.
Now, again due to the finiteness of $S$ there exist $k,m$ with 
$1\leq k < m \leq |S|+1$ and $s_k=s_m$.
Hence we can define 
$x\eqdef v^1_{1,m-1}av^2_{1,m-1}a\cdots av^k_{1,m-1}a$,
$v\eqdef v^{k+1}_{1,m}av^{k+2}_{1,m}a\cdots av^m_{1,m}a$,
$y\eqdef v^{k+1}_{1,m-1}av^{k+2}_{1,m-1}a\cdots av^{m-1}_{1,m-1}au_{1,m-1}$, and
$z\eqdef w^{''}_r$, where $r$ is the index such that $w_{1,m-1}=w^{'}_r$.
Note that $xy=w^{'}_r$. 
We conclude with the assertions of the claim, that
\[ 
\renewcommand{\arraystretch}{1.5}
ya  \begin{array}[t]{cl}
      =      &  v^{k+1}_{1,m-1}a\cdots av^{m-1}_{1,m-1}au_{1,m-1}a \\
      \sw    &  v^{k+1}_{\tau(1),m-1}a\cdots av^{m-1}_{\tau(1),m-1}av^m_{1,m}a \\
      =      &  v^{k+1}_{1,m}a\cdots av^{m-1}_{1,m}av^m_{1,m}a \\
      =      &  v
          \end{array}
\]
Moreover we see that $\delta(s_0,xyz)=\delta(s_0,w^{'}_rw^{''}_r)=\delta(s_0,w_{2r-1})\not\in S^{'}$
and $\delta(s_0,xyaz)=\delta(s_0,w^{'}_raw^{''}_r)=\delta(s_0,w_{2r})\in S^{'}$.
This shows that $F$ has pattern P$_1$.

{\em Case 2}.
Now assume that there exists an infinite strictly decreasing subsequence of $\{\alpha_i\}$.
Then obviously $\{w^R_j\}$ is an infinite alternating extension chain with respect to
$L(F)^R$. Let $\hat{F}$ be a dfa accepting $L(F)^R$.
Attaching markers $\alpha^{'}_i$ in the same way as above leads to $\alpha^{'}_i=1-\alpha_i$ and
hence there is a strictly increasing subsequence of $\{\alpha^{'}_i\}$.
We can conclude as in case 1 that $\hat{F}$ has pattern P$_1$.
This finishes the proof of $(2)\Rightarrow (1)$ and 
we turn to the remaining implications.

$(1)\Rightarrow (4)$:
Suppose some dfa $F$ has pattern P$_3$.
Then we have for $i\geq 0$ the infinite alternating word extension chain
$xv^iyzu^iz^{'}\sw xv^iyazu^iz^{'} \sw xv^{i+1}yzu^{i+1}z^{'}$ since
either $ya\sw v$ or $az\sw u$.

$(4)\Rightarrow (3)$: 
If some dfa $F$ has pattern P$_2$ then this is also
a pattern P$_3$ (with $v=y=\eps$), which is a contradiction.
Next we show that if some dfa $\hat{F}$ has pattern P$_2$
then any dfa $F$ with $L(F)=L(\hat{F})^R$ has pattern P$_1$, 
and again this is also pattern P$_3$ (with $u=z^{'}=\eps$), a contradiction as well.
So suppose that a dfa $\hat{F}=(A, \hat{S}, \hat{\delta}, \hat{s}_0, \hat{S}^{'})$ 
has the pattern P$_2$ witnessed by $x,z,u,z^{'}\in A^*$ and $a\in A$.
Let $F=(A, S, \delta, s_0, S^{'})$ be any dfa with $L(F)=L(\hat{F})^R$ and
choose $m,k\in\N$ with $m>k\geq 0$ such that 
$\delta\big(s_0,(z^{'})^R(u^R)^k\big)=\delta\big(s_0,(z^{'})^R(u^R)^{k+m}\big)$.
We define
$\bar{x}\eqdef (z^{'})^R(u^R)^k$, $\bar{v}\eqdef (u^R)^m$, 
$\bar{y}\eqdef z^R$ and $\bar{z}\eqdef x^R$.
Now one can easily verify that $\bar{x}, \bar{v}, \bar{y}, \bar{z}\in A^*$ and $a\in A$
give rise to  pattern P$_1$ in $F$ since $\bar{y}a\sw\bar{v}$ follows from $az\sw u$.

$(3)\Rightarrow (2)$.
Suppose that a dfa $F=(A, S, \delta, s_0, S^{'})$ has pattern P$_1$
witnessed by $x,v,y,z\in A^*$ and $a\in A$.
Let $\hat{F}=(A, \hat{S}, \hat{\delta}, \hat{s}_0, \hat{S}^{'})$ be any dfa with 
$L(\hat{F})=L(F)^R$ and 
choose $m,k\in\N$ with $m>k\geq 0$ such that  
$\hat{\delta}\big(\hat{s}_0,(yz)^R(v^R)^k\big)=\hat{\delta}\big(\hat{s}_0,(yz)^R(v^R)^{k+m}\big)$ and
$\hat{\delta}\big(\hat{s}_0,(yaz)^R(v^R)^k\big)=\hat{\delta}\big(\hat{s}_0,(yaz)^R(v^R)^{k+m}\big)$.
We define
$\bar{x}\eqdef z^R$, $\bar{u}\eqdef (v^R)^m$, 
$\bar{z}\eqdef y^R(v^R)^k$ and $\bar{z}^{'}\eqdef x^R$.
Again, one can easily verify that $\bar{x}, \bar{u}, \bar{z}, \bar{z}^{'}\in A^*$ and $a\in A$
give rise to pattern P$_2$ in $\hat{F}$ since $a\bar{z}\sw\bar{u}$ follows from $ya\sw v$.
\endproof

We remark that the proof of $(2)\Rightarrow (1)$ even shows that 
the automata $F$ and $\hat{F}$ do not have the {\em two} instances of pattern P$_1$ with
$s\in S^{'}$ and $s^{'}\not\in S^{'}$ on one hand, 
and $s^{'}\in S^{'}$ and $s\not\in S^{'}$ on the other hand.
The same holds analogously for the other patterns.
To see this note that we can start the whole investigation at the very beginning of the
proof with the sequence $\{w_{j+1}\}$.

Using the above Theorem we obtain a co-NL(=NL)-algorithm for 
the decision problem for $\L{1}$ simply by testing the
occurence of the pattern P$_3$ in a given dfa. This algorithm
is completely different from those which
follow from the characterizations in \cite{simon75, stern85}.
Note that S.~Cho and D.T.~Huynh proved in \cite{ch91}
that the decision problem for $\L{1}$ is even NL-complete.

%

\section{Complexity Theoretical Consequences}

Let a nondeterministic polynomial time Turing machine $M$ output on every path a 
symbol from $A$ and assume a fixed ordering on the set of all paths.
We additionally assume here that, given some input $x$ and the number of a 
path $i$, one can compute in polyomial time the output of $M$ on path $i$
({\em balanced} computation tree).
This leads in a natural way to the notion of the {\em leafstring} of $M$ on some input $x$ 
when concatenating the output symbols of $M$'s computation tree.
Now a language $L\seq A^*$ gives rise to the class $\leafp(L)$ of all languages $L'$ 
for which there is
a machine $M$ of the above type such that for all $x$ it holds that $x\in L'$ if and only if
the leafstring of $M$ on input $x$ belongs to $L$.
Furthermore, for some class $\C$, denote by $\leafp(\C)$ the union of
all classes $\leafp(L)$ with $L\in \C$.

As stated in the introduction this leaf language approach led to new insights into
the structure of complexity classes between $\P$ and ${\rm PSPACE}$.
However, most results deal with {\em classes} of leaf languages and an important
question is what complexity classes are definable by a {\em single} leaf language.
Some progress in this direction has been made in \cite{bo95, bokust98}.

Due to the close connection of the classes of the Straubing-Th\'erien hierarchy
to ${\rm FO}[<]$-logic (Theorem \ref{logic})
we can make use of the known relationship between languages definable within this 
logic and the classes of the polynomial time hierarchy.

\begin{thm}[\cite{buvo98}] \label{buvo}
Let $A$ be an arbitrary alphabet with $|A|\geq 2$ and let $k\geq 1$.
\begin{enumerate}
\item $ \Sigp{k} =  \leafp(\L{k-1/2}) $
\item $ \Pip{k} =   \leafp({\rm co}\L{k-1/2})$
\end{enumerate}
\end{thm}

The ``forbidden-pattern'' characterization of the classes $\L{1/2}$ 
from \cite{pinweil97} enables us to show which complexity classes
are exactly definable by a single leaf language from this class. 
 
\begin{thm}
For an arbitrary alphabet $A$ with $|A|\geq 2$ we have 
\[
   \set{\leafp(L)}{L\in\L{1/2}} = \big\{ \{\emptyset\},
                                         \{B^*~|~B~{\rm finite~alphabet}\},
                                         \P, \NP
                                  \big\}
\]
and given some dfa accepting a language $L\in\L{1/2}$ one can effectively
determine the class on the right hand side with which $\leafp(L)$ coincides. 
\end{thm}

For single leaf languages from the boolean hierarchy over $\L{1/2}$ the situation
is a lot more complicated.
However, we have the following ``union-style'' theorem which provides an upper bound 
for complexity classes definable via such leaf languages.
Throughout the paper we studied the classes $\L{1/2}(k)$ for an arbitrary but fixed
alphabet $A$. Now we will emphasize on the chosen alphabet and denote 
by $\L{1/2}^A(k)$ the classes $\L{1/2}(k)$ defined for languages over $A$.

\begin{thm}
For any $k\geq 1$,
\[
   \NP(k)~~=  \bigcup_{A~{\rm finite~alphabet}} \leafp\big(\L{1/2}^A(k)\big).
\]

\end{thm}

\proof
To see the inclusion from right to left note with Theorem \ref{buvo}.1 that
$\leafp(\L{1/2}^A)\seq\NP$ for any alphabet $A$.
Furthermore it holds for languages $L_1, L_2$ that 
$\leafp(L_1\cup L_2)\seq\leafp(L_1)\vee\leafp(L_2)$,
$\leafp(L_1\cap L_2)\seq\leafp(L_1)\wedge\leafp(L_2)$ and
$\leafp(\overline{L_1})=\co\leafp(L_1)$, where
$\C_1\vee\C_2\eqdef\set{L'_1\cup L'_2}{L'_1\in\C_1, L'_2\in\C_2}$ and
$\C_1\wedge\C_2\eqdef\set{L'_1\cap L'_2}{L'_1\in\C_1, L'_2\in\C_2}$
for classes $\C_1, \C_2$.

For the other inclusion define for $k\geq 1$ the alphabet $A_k\eqdef\{0,1,2,\ldots,k\}$
and the language $L_k\eqdef\set{w\in A_k^*}{ \max\{i\in A_k~|~i\sw w\}~{\rm is~odd}}$.
One can show with Lemma \ref{nf} that $\leafp(L_k)=\NP(k)$.
Observe that $m^+(L_k)=k-1$, so 
with Theorem \ref{Sk-mk} it follows that $L_k\in\L{1/2}^{A_k}(k)$.
\endproof 

\begin{coro}
If $m^+(L)<k$ for a regular language $L$ then $\leafp(L)\seq\NP(k)$. 
\end{coro}

Note that the measure $m^+$ is computable (Theorem \ref{compm}).
Moreover the results obtained here remain valid if we omit the restriction
that the computation tree of a Turing machine must be balanced. 

Finally we compare our results with related work.
In \cite{crhevowa98} the case of commutative leaf languages has been studied, 
i.e.~the case where membership to a language depends only on the numbers of occurences
of the alphabet symbols.
For an oracle $D$
we denote by $\C^D$ the relativized version of a complexity class $\C$.
It has been proved in the mentioned paper that for every commutative language $L$,
\[
 m^+(L)<k \Longleftrightarrow \forall D \big(\leafp(L)^D \seq \NP(k)^D\big).
\]
Furthermore, other (stronger) measures $n^+$ and $n^-$ have been defined,
i.e.~$n^+(L)\leq m^+(L)$ and $n^-(L)\leq m^-(L)$,
and it has been proved that for every commutative language $L$,
\[
 n^-(L)\geq k \Longleftrightarrow \forall D \big(\leafp(L)^D \supseteq \NP(k)^D\big).
\]

%

\bibliographystyle{alpha}
\bibliography{pt}

\end{document}